\documentclass[journal,12pt,onecolumn]{IEEEtran}

\usepackage{isit13_ownstyle}
 \usepackage[utf8]{inputenc}
\author{Abolfazl Motahari, Kannan Ramchandran, David Tse and Nan Ma\\
Department of Electrical Engineering and Computer Sciences\\
University of California, Berkeley\\
\{motahari,kannan,dtse\}@eecs.berkeley.edu}
\begin{document}
\title{Optimal DNA shotgun sequencing:\\ Noisy reads are as good as noiseless reads}

\maketitle

\begin{abstract}
We establish the fundamental limits of DNA shotgun sequencing under noisy reads. We show a surprising result: for the i.i.d. DNA model, noisy reads are as good as noiseless reads, provided that the noise level is below a certain threshold which can be  surprisingly high. As an example, for a uniformly distributed DNA sequence and a symmetric substitution noisy read channel, the threshold is as high as 19\%.
\end{abstract}


\section{Introduction}

\begin{figure}
\begin{center}
\includegraphics[width=7in]{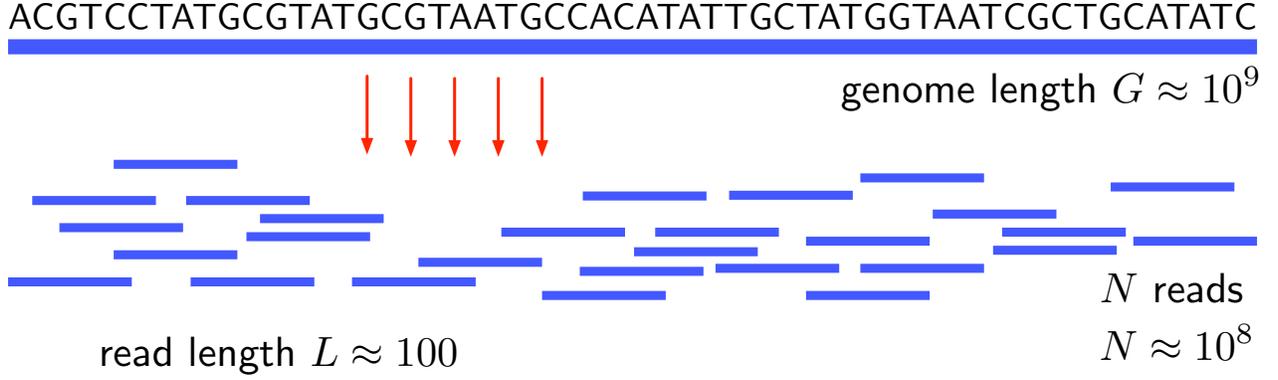}
\end{center}\vspace{-0.5cm}
\caption{Schematic for shotgun sequencing.}
\label{fig:setup}
\end{figure}

DNA sequencing is the basic workhorse of modern day biology and medicine. Since the sequencing of the Human Reference Genome ten years ago, there has been an explosive advance in sequencing technology. Multiple ``next-generation" sequencing platforms  have emerged.  All of them are based on the whole-genome shotgun sequencing method.
The basic shotgun DNA sequencing set-up is shown in Figure~\ref{fig:setup}. Starting with a DNA molecule, the goal is to obtain the sequence of bases ($A,C,G$ or $T$) comprising it.  The sequencing machine extracts a large number of reads from the DNA; each read is a randomly located fragment of the DNA sequence. The {\em DNA assembly problem} is to reconstruct the DNA sequence from the many reads.

A basic question, still largely open, is the following: given DNA sequence statistics and characteristics of the sequencing technology such as read length and noise statistics, how many reads are needed to reconstruct the original DNA sequence, if it is possible at all? The answer to this question can provide an algorithm-independent basis for evaluating the efficiency of a sequencing technology and can be used to compare different assembly algorithms.
\cite{MBT12} provides an answer to this question in a simple setting: 1) each read has the same length $L$ bases  and is uniformly and independently sampled from the length $G$ DNA sequence; 2)the DNA sequence is modeled as an i.i.d.  string; 3) the read process is noiseless.
The main result shows that in the asymptotic regime where $L$ and $G \rightarrow \infty$ with  $ \bar{L} = L/\log G$
fixed, a critical phenomenon occurs: when $\bar{L} <  \Lcrit $, reconstruction is impossible, and when $\bar{L} > \Lcrit$, then having enough reads to {\em cover} the DNA sequence is also sufficient for reconstruction.  Here, $\Lcrit = 2/\Hr$, where $\Hr$ is the Renyi entropy rate of order  $2$.  The significance of $\Lcrit$ is that with high probability, there are no repeats of length more than $\Lcrit$ in the DNA sequence. The coverage bound  is a well-known lower bound introduced by Lander and Waterman \cite{LW88} in the early days of sequencing. Thus, the result says that as long as the read length is longer than the longest repeat in the DNA, this  lower bound is asymptotically tight.

In \cite{BBT13}, the theory of noiseless assembly is extended to DNA sequences with arbitrary repeat statistics. In this paper, instead, we keep the i.i.d. DNA model but we consider {\em noisy} reads.

The optimal assembly algorithm which achieves the fundamental limit in the above setting is the {\em  greedy algorithm}. The greedy algorithm merges reads with the largest overlap first, where the overlap between two reads is the longest exact match between a prefix of one read and a suffix of another read. A natural extension of the greedy algorithm to the noisy read case is that instead of looking at exact matches, one allows {\em approximate} matches, where the degree of approximation tolerated is a function of the read noise statistics. The performance analysis of such an algorithm under noisy reads was considered in \cite{MBT12}.  Not so surprisingly, noise {\em always} degrades the performance of the greedy algorithm, and in fact the effect is quite significant.

The modification of the greedy algorithm is only one approach to deal with noise. But are there better approaches? What, in fact, is the fundamental limit on the system performance under noisy reads? We show  a surprising result in this paper: provided that the noise level is below a certain threshold, noise has {\em no } impact on the asymptotic performance.  The threshold on the noise level is given by the condition
\begin{equation}
 \label{eq:threshold}
 I_{\rm read} > \Hr,
 \end{equation}
where 
$$I_{\rm read}= \min_{s \in \{A,G,C,T\}} I(S=s; Y),$$
with:
$$ I(S=s;Y) :=  \sum_y \pi(y|s) \log \left(\frac{ \pi(y|s)}{\sum_s Q_s \pi(y|s)}\right).$$
Here, $Q_s$ is the probability that a DNA base equals $s$ and $\pi(y|s)$ is the probability that a DNA base $s$ is read as $y$ through the noisy read channel. In particular, under the uniform distribution $Q_s = 0.25$ for all $s$ and symmetric read channel with probability of mis-read $\delta$, $\Hr = 2$ bits and $I_{\rm read}$ is the capacity of the read channel:
$$ I_{\rm read}= - \delta \log \frac{\delta}{3} - (1-\delta) \log (1-\delta).$$
The condition (\ref{eq:threshold}) translates to a threshold of $\delta^* = 0.19$ for this example. As long as the noise level is below $19\%$, the noiseless performance can be achieved, i.e. coverage is sufficient when $\bar L > \Lcrit = 2/\Hr$.

In communication, noise almost always has a detrimental effect on asymptotic performance, as it degrades the channel capacity. So we would like to give some intuition on why noise (below a certain level) has no impact on the asymptotic performance in the shotgun assembly problem considered here. First, we need to understand better the implication of the coverage condition. It follows from Lander-Waterman's results that the number of reads $\Ncov$ needed to cover the entire DNA sequence with probability at least $1-\epsilon$ is well approximated by:
$$ \Ncov  \approx \frac{G}{L} \ln \left (\frac{G}{L\eps} \right ).$$
Thus, the coverage depth, i.e. the average number of reads covering each base, is given by:
$$ c := \frac{\Ncov L}{G} \approx \ln \left (\frac{G}{L\eps} \right ).$$
For example, for $G = 3 \times 10^9$, $L= 100$, $\epsilon = 0.05$, the coverage depth $c = 20$.
In the asymptotic limit, the coverage depth goes to infinity. This high coverage depth provides a level of redundancy which can be exploited to deal with noisy reads: if multiple reads covering the same region of the DNA can be aligned together, then one can average over the symbols in the different noisy reads to obtained a {\em cleaned-up} read. However, if the read length is too short, this alignment cannot be done accurately, since noisy reads from other similar-looking regions of the DNA will be mis-aligned together and this would not help the noise averaging process. This minimum read length for accurate alignment would depend both on the noise statistics and the repeat statistics of the DNA sequence. What we show is that for the i.i.d. DNA model and memoryless read noise,  as long as the noise level is less than the threshold given by condition (\ref{eq:threshold}), then accurate alignment can be achieved provided that the read length is longer than the longest repeat on the DNA sequence. This is exactly the same condition on the read length needed for noiseless assembly.  Hence, one essentially can achieve error correction for free.


The scheme we propose to achieve the fundamental limit under noisy reads has two stages: an {\em error-correction} phase, which aligns reads from the same region of the DNA and averages across them to produce cleaner reads, followed by an {\em assembly} phase, applying the greedy algorithm with approximate match to the cleaner reads. Provided that the noise level satisfies condition (\ref{eq:threshold}) to allow accurate read alignment, the noise level of the reads can be driven to be vanishingly small after the error-correction phase. Since it was shown in \cite{MBT12} that the performance of the greedy algorithm is continuous in the noise level, this implies noiseless performance can be achieved asymptotically.

In the assembly literature, there are two approaches to deal with the noise in the reads. In the first approach,  error-correction is performed jointly with assembly such as Velvet \cite{velvet} and ABySS \cite{ABySS} which are based on de Bruijn graph. In the second approach, error-correction is performed first, followed by an assembly algorithm which assumes the reads are essentially clean. Examples of the algorithms are SHREC \cite{Shrec}, Reptile \cite{reptile}, and Quake \cite{quake}. The latter is a {\em separation} approach, which is conceptually simpler. What we show in this paper is that, at least for the simple model considered here, the separation approach is in fact information-theoretically optimal, up to a certain threshold on the noise level.


\section{Formulation and Previous Results}
\label{sec:formulation}

\subsection{DNA Model}

The DNA sequence  $\bs=s_1 s_2 \dots s_G$ is modeled as an i.i.d. random string of length $G$ with each symbol taking values according to a probability distribution $Q=(Q_A,Q_C,Q_G,Q_T)$ on the alphabet $\{A,C,G,T\}$. To avoid boundary effects, we assume that the DNA sequence is circular, i.e., $s_i=s_j$ if $i=j$ mod $G$; this simplifies the exposition, and all results apply with appropriate minor modification to the non-circular case as well.

\subsection{Noiseless Reads}
A noiseless read is a substring of length $L$ from the DNA sequence. The set of reads is denoted by $\RR =\{\br_1,\br_2,\ldots,\br_{N}\}$. The starting location of read $i$ is $t_i$, so $r_i=\bs[t_i,t_i+L-1]$.  The set of starting locations of the reads is denoted $\TT=\{t_1,t_2,\ldots,t_{N}\}$, where we assume $1\leq t_1\leq t_2\leq \dots\leq t_{N}\leq G$. We assume that the starting location of each read is uniformly distributed on the DNA and the locations are independent from one read to another.

An \emph{assembly algorithm} takes a set of $N$ reads $\RR=\{\br_{1},\dots,\br_N\}$ and returns an estimated sequence $\hat \bs=\hat\bs(\RR)$.
We require \emph{perfect reconstruction}, which presumes that the algorithm  makes an error if $\hat\bs\neq \bs$.  A question of central interest is: what are the conditions on the read length $L$  and the number of reads $N$ such that the reconstruction error probability is less than a given target $\epsilon$ for some algorithm?  Define  the {\em minimum normalized coverage depth} $\capacity (\bar L)$:
\begin{equation}
\capacity (\bar L ) = \lim_{G \to \infty, L=\bar L \log G} \frac{\Nmin(\epsilon,G,L)}{\Ncov(\epsilon,G,L)},
\end{equation}
where $\Nmin(\epsilon,G,L)$ is the minimum number of reads required to reconstruct the DNA sequence with probability at least $1-\epsilon$ and $\Ncov(\epsilon,G,L)$ is the minimum number of reads to cover the DNA sequence with probability at least $1-\epsilon$.

The main result for this noiseless read model is:
\begin{theorem}\label{thm:noiseless_assembly}
\cite{MBT12} Fix an $\epsilon < 1/2$. $\capacity(\bar L)$ is given by
\begin{equation}
\label{eq:capacity}
\capacity(\bar L)=
\begin{cases}
\infty & \text{if} ~ \bar{L}  < 2/\Hr,\\
1 & \text{if} ~ \bar{L} > 2/\Hr,
\end{cases}
\end{equation}
where $\Hr$ is the Renyi entropy of order 2 defined as:
\begin{equation}
\label{eq:Renyi}
\Hr := - \log \sum_{s\in\{A,C,G,T\}} Q_s^2.
\end{equation}
\end{theorem}

\subsection{Noisy Reads}
\label{sec:formulation-noisy}
Now we assume that the read process is noisy and consider a simple probabilistic model for the noise. A base $s\in\{A,C,G,T\}$ is read to be $y\in \YY$ for some ground set $\YY$ with probability $\pi(y|s)$. Each base is perturbed independently, i.e. if $\br=r_1,\dots,r_L$ is a read from the physical underlying subsequence $\bs=s_1,\dots,s_L$ of the DNA sequence, then
$$\P(\br | \bs) =\prod_{i=1}^L \pi(r_i| s_i).$$
Moreover, it is assumed that the noise affecting different reads is independent.

In the noiseless read case, we aim for  {\em perfect reconstruction}. In the noisy read case, we aim for {\em perfect layout}. By perfect layout, we mean that all the reads are mapped correctly to their true locations. Note that perfect layout does not imply perfect reconstruction as the consensus sequence may not be identical to the DNA sequence on every single base. On the other hand, since coverage implies that most positions on the DNA are covered by many reads (growing with $G$), the consensus sequence will be correct in most positions if we achieve perfect layout.

By modifying the greedy algorithm to allow for approximate instead of exact matches, the following performance can be achieved.

\begin{figure}
\begin{center}
\includegraphics[width=5in, height=3.5in]{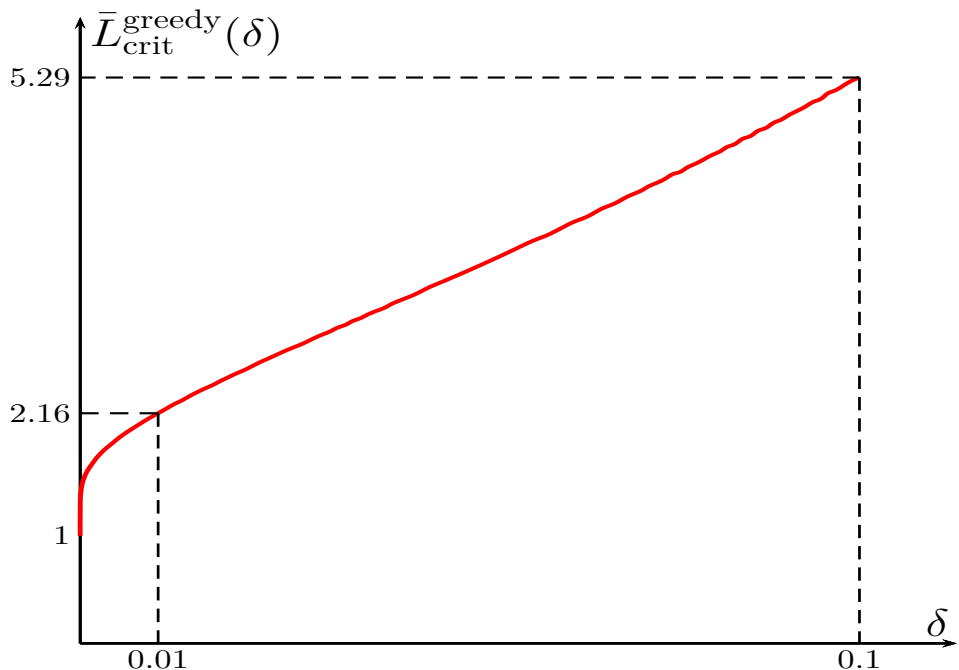}
\end{center}
\caption{Plot of $\Lcg(\delta)$  as a function of the noise level ~$\delta$ for the uniform source and  symmetric noise model.}
\label{fig:I*}
\end{figure}

\begin{theorem}\label{thm:approx_greedy}
 \cite{MBT12} The modified greedy algorithm can achieve normalized coverage depth $c(\bar L) =1$ if $\bar{L}> \Lcg$. $\Lcg$ is a continuous function of the DNA and noise statistics and is strictly larger than $\Lcrit$ whenever the noise is non-trivial.
\end{theorem}
Fig. \ref{fig:I*} gives an example of $\Lcg$.


\section{Optimal Error Correction}
\label{sec:noisy}

Theorem \ref{thm:approx_greedy} shows that the critical read length increases from that in the noiseless case when the modified greedy algorithm is {\em directly} applied on the noisy read data. What we show in this section is that, if the noise level is below a certain threshold, there is actually enough redundancy  in the noisy reads to perform almost perfect error correction. By applying the modified greedy algorithm on the cleaned-up reads, noiseless performance can be achieved asymptotically.

First, we define the quality of a cleaned-up read $\tbr$ of length $K$:
$$ d(\tbr) = \min_{\bx} \frac{d_{H}(\tbr,\bx)}{K},$$
where the minimization is over all length $K$ subsequences of the DNA sequence $\bs$. Also, let $\pbr (\tbr)$ be the minimizing subsequence, i.e. the one on the DNA sequence with the closest match to $\tbr$.

The main result of the paper is the following theorem.

\begin{theorem}
\label{thm:main}
(Error Correction) Assume: $\L > \Lcrit$, the noisy reads cover the DNA sequence asymptotically, and the noisy read channel satisfies condition \ref{eq:threshold}.
Then there is an error-correction algorithm which takes as inputs the $N$  noisy reads $\br_1, \ldots, \br_N$ and outputs $\tilde{N}$  cleaned-up reads $\tbr_1, \ldots,\tbr_{\tilde{N}}$ such that:

\begin{itemize}
\item Each cleaned-up read $\tbr_i$ is of length $K$ such that $K/L \rightarrow 1$.

\item There is a sequence $\{\tau_G\}$ with $\tau_G \rightarrow 0$ such that:
$$ \lim_{G \rightarrow \infty} \P(\max_i d(\tbr_i) > \tau_G)  = 0.$$

\item Coverage: $\pbr(\tbr_1), \ldots, \pbr(\tbr_{\tilde{N}})$ cover the DNA sequence asymptotically.

\end{itemize}

\end{theorem}

\section{Proof of Theorem \ref{thm:main}}

The proof technique is based on the method of types \cite{csiszarinformation} and the slight modification of strong typicality, as  defined in \cite{Orlitsky2001}. Let $\XX$ be discrete set of size $|\XX|$. The set of all possible probability distributions on $\XX$ is denoted by $\PP(\XX)$. The set of all possible emprical distributions (types) of sequences $\bx \in \XX^K$ is denoted by $\PP_K(\XX)$. Clearly, $\PP_K(\XX) \subset \PP(\XX)$. The cardinality of $\PP_K(\XX)$ is upper bounded by $(K+1)^{|\XX|}$ \cite{csiszarinformation}.

We say a sequence $\bx\in \XX^K$ is typical wrt the probability distribution $F$, if
$$ |  \frac{N(a|\bx)}{K}- F(a) | \leq \epsilon F(a),$$
for all $a\in \XX$. Here, $N(a| \bx)$ is the number of occurrences of $a\in \XX$ in $\bx$. Similarly, one can define the joint typicality of a set of sequences $\bX := \{\bx_1,\bx_2, \ldots, \bx_n\}$ wrt the probability distribution F on $\XX^n$.  The main property of this definition of joint typicality is that if  a set $\bX$ of sequences is jointly typical wrt F, any subset $\SS$ of the sequences is jointly typical wrt the marginal distribution of F on $\SS$.

\subsection{The Error Correction Algorithm}
Let $\PP_K(\{A,C,G,T\})$ denote the set of all possible types of sequences in $\{A,C,G,T\}^K$. 
For given $P\in \PP_K$, we denote $F_P^M$ to be the distribution of observing $M$ independent samples of a base through the noisy read channel with the base distribution $P$. Clearly,
\begin{equation}
F_P^M(y_1,\dots,y_M): = \sum_{s\in \{A,C,G,T\}} \left(\prod_{i=1}^M \pi(y_i | s)\right) P(s).
\end{equation}
We also denote $F_P^1$ by $F_P$.

Let $\RR$ be the set of all reads. For a fixed $K$ and for each read, we extract all the substrings of length $K$ from that read where each substring is called a $K$-mer. We create the pool $\UU$ consisting of all the $K$-mers. 

A set of $M$ $K$-mers, $\bU= \{\bu_1, \dots,\bu_M\}$ is said to be a {\em good alignment} if $\bU$ is jointly typical wrt $F_P^M$ for some $P\in \PP_K$. In the definition of ``good alignment'', we have considered all possible empirical distributions over DNA bases instead of considering only $Q$  which is the true distribution of DNA bases. The reason is that the DNA sequence is long and contains atypical sequences of length $K$ wrt the true distribution $Q$. Therefore, if we only use $Q$ to define ``good alignment'', we lose the coverage of the DNA sequence.

For any good alignment $\bU$, we take, for each component $i=1, \ldots, K$, the Maximum Likelihood (ML) estimate $\hat{s}_i$ of the underlying base $s_i$ from $u_{1i}, \ldots, u_{Mi}$, assuming they are independent observations of $s_i$ through the read channel. The averaged $K$-mer $(\hat{s}_1, \ldots, \hat{s}_K)$ is denoted by $\tbr_{\bU}$. Lastly, we create $\tilde{\RR}$, the list of all cleaned-up reads, consisting of the averaged $K$-mers from all possible good alignments taken from $\UU$.

\subsection{Analysis of the Algorithm}

To analyze the proposed error correction algorithm, we set $K=L-L^{\alpha}$ and $M=\beta \log L$ for some constant $\alpha$ and $\beta\in (0,0.5)$.

\subsubsection{Error Correction Condition}
%
%

To show the error correction condition of Theorem \ref{thm:main}, we define $\EE$ be the event that there is a  good alignment from $\UU$ such that the averaged sequence $\tbr$ has quality $d(\tbr) > \tau_G$.  We define the following events:
\noindent
\begin{description}
\item[\!\!\!\!\!$\EE_1(J)\! :$] There is a good alignment from $\UU$ with  $J<M$ $K$-mers  from distinct locations of the DNA sequence.
\item[\!\!\!\!\!$\EE_2(M_1)\! :$] There is a good alignment from $\UU$ with two subsets of $K$-mers each of which coming from the same location  and having size $M_1< \frac{M}{2}$.
\item[\!\!\!\!\!$\EE_3(M_0)\!:$] There are $M$ K-mers in $\UU$  with at least $M_0$ $K$-mers from a single location but whose averaged sequence  is not within Hamming distance $K\tau_G$ from the DNA subsequence at that location.
\end{description}
We claim that $\EE\subseteq \EE_1(J) \cup \EE_2(M_1) \cup \EE_3(M-(M_1-1)(J-2))$.  This is due to the fact that if a good alignment has less than $J$ $K$-mers coming from distinct locations and has at most one subset of $K$-mers  of size $M_1$ coming from the same location, then at least $M-(M_1-1)(J-2)$ $K$-mers come from a single location.

Therefore, the union bound gives us
\begin{equation}\label{eq:unionB}
\P(\EE) \leq \P(\EE_1(J)) + \P(\EE_2(M_1)) + \P(\EE_3(M-(M_1-1)(J-2))).
\end{equation}
In particular, we will set $J = M^{\frac{1}{4}}+2, M_1 = M^{\frac{1}{4}} +1$, and hence $M - (M_1-1)(J-2) = M - \sqrt{M}$, i.e., we are interested in the case when the majority of the reads come from a single location in event $\EE_3$.  We will upper bound each of the terms in \eqref{eq:unionB} and show that they all go to zero as $M \rightarrow \infty$.

\paragraph{$\P(\EE_1(J))\rightarrow 0$}
The event $\EE_1(J)$ happens when there  is a good alignment $\bU$ containing $J$ $K$-mers from distinct locations.
Without loss of generality, let us assume that the first $J$ $K$-mers of $\bU=\{\bu_1,\dots,\bu_M\}$ are sampled from distinct locations. Since $\bU$ is jointly typical wrt $F_P^M$ for some $P\in\PP_K$, $\{\bu_1, \ldots \bu_J\}$ is jointly typical wrt $F_P^J$.
Considering the fact that there are at most $G^J$ possible choices for $J$ distinct locations on the DNA sequence and there are only polynomially many types, we can apply large deviation arguments to obtain
\begin{equation}\label{eq:Ej}
\P(\EE_1(J)) \leq (K+1)^4 G^J 2^{- K \left( \min_{P\in \PP_K} D\left( F_P^J || \prod_{i=1}^J F_{Q} \right)-\delta_1(\epsilon)\right)},
\end{equation}
for some $\delta_1(\epsilon)>0$.
We proceed by computing the KL-divergence $D\left(F_P^J || \prod_{i=1}^J F_{Q} \right)$. Let $S$ denote a DNA base distributed according to $P$ and $Y_1, \ldots Y_J$ be independent observations of $S$ through the read channel. Then,
\begin{align*}
D\left(F_P^J || \prod_{i=1}^J F_{Q}^1 \right) & = \sum_{(y_1,\dots,y_J)\in \YY^J} F_P^J(y_1,\dots,y_J) \log\left(\frac{F_P^J(y_1,\dots,y_J)}{\prod_{i=1}^J F_{Q}(y_i)} \right)  \\
& = \sum_{(y_1,\dots,y_J)\in \YY^J} F_P^J(y_1,\dots,y_J) \log\left(\frac{1}{\prod_{i=1}^J F_{Q}(y_i)} \right)-H(Y_1,\dots,Y_J)  \\
& = J \sum_{y\in \YY} F_P(y) \log\left(\frac{1}{ F_{Q}(y)} \right)-H(Y_1,\dots,Y_J)  \\
& \stackrel{(a)}{\geq} J \sum_{y\in \YY} F_P(y) \log\left(\frac{1}{ F_{Q}(y)} \right) - H(Y_1,Y_2,\dots,Y_J,S) \\
& = J \sum_{y\in \YY \atop s\in\{A,C,G,T\}} P(s) \pi(y|s) \log\left(\frac{\pi(y|s)}{ F_{Q}(y)} \right) - H(S)\\
& \geq J \sum_{y\in \YY \atop s\in\{A,C,G,T\}} P(s) \pi(y|s) \log\left(\frac{\pi(y|s)}{ F_{Q}(y)} \right) - 2,
\end{align*}
where $(a)$ comes from the fact that entropy increases by adding a new random variable. Next, we need to minimize the obove expression over all distributions in $\PP_K$. However, we obtain slightly looser bound if we minimize over all distribution in $\PP$. Let $P^* \in \PP$ be the minimizer of the following program
\begin{align*}
E_1 = \min_{P\in \PP} J \sum_{y\in \YY \atop s\in\{A,C,G,T\}} P(s) \pi(y|s) \log\left(\frac{\pi(y|s)}{ F_{Q}(y)} \right) - 2.
\end{align*}
The optimization problem is a linear program and an optimal $P^*$ can be obtained by choosing a letter $s^*$ and put all the probability mass on it, with $s^*$ given by:
\begin{equation}
s^* = \text{arg} \min_s  \sum_x \pi(y|s) \log \left(\frac{ \pi(y|s)}{F_{Q}(y)}\right)
\end{equation}
and hence $E_1$ becomes
\begin{align*}
E_1 & = J \sum_y \pi(y|s^*) \log\left(\frac{\pi(y|s)}{ F_{Q}(y)} \right) - 2\\
 &= JD( \pi(y|s^*) || F_{Q})-2.
\end{align*}
Hence, since $\L > 2/\Hr$, provided that  condition (\ref{eq:threshold}) is satisfied, $\P(\EE_1(J)) \rightarrow 0$ as $J,G \rightarrow \infty$.

\paragraph{$\P(\EE_2(M_1)) \rightarrow 0$}
The event $\EE_1(J)$ happens when there  is a good alignment $\bU$ containing two sets of size $M_1$ each sampled from a single location on the DNA sequence.
Without loss of generality, we assume that the first $M_1$ $K$-mers are sampled from position $k_1$ on the DNA sequence and the second $M_1$ $K$-mers are sampled from position $k_2$ on the DNA sequence with $k_1\neq k_2$. Both of these sets of $K$-mers are jointly typical wrt $F_P^{M_1}$ for some $P\in \PP_K$.
Considering the fact that there are at most $G^2$ possible choices for $k_1$ and $k_2$ and applying large deviation arguments, we obtain
\begin{equation}\label{eq:e2}
\P(\EE_2) \leq (K+1)^4 G^2 2^{- K \left(\min_{P\in\PP_K} D\left(F_P^{2M_1} || F_{Q}^{M_1}F_{Q}^{M_1} \right)-\delta_2(\epsilon)\right)},
\end{equation}
for some $\delta_2(\epsilon)$.
We proceed by computing the KL-divergence:
\begin{equation}
D\left(F_P^{2M_1} || F_{Q}^{M_1}F_{Q}^{M_1} \right) =D\left(F_P^{2M_1} || F_{P}^{M_1}F_{P}^{M_1} \right)+2 D\left(F_P^{M_1} || F_{Q}^{M_1} \right).
\end{equation}
Let $\hat s_1$ and $\hat s_2$ be the ML estimate of underlying base $s$ obtained from  $y_1,\dots,y_{M_1}$ and $y_{M_1+1},\dots,y_{2M_1}$, respectively. The data processing inequality, c.f. \cite{csiszarinformation}, implies
\begin{equation}
D\left(F_P^{2M_1} || F_{Q}^{M_1}F_{Q}^{M_1} \right) \geq D\left(P(\hat{s}_1,\hat{s}_2) || P(\hat{s}_1)P(\hat{s}_2)\right)+2 D\left(P(\hat{s}_1) || Q(\hat{s}_1) \right).
\end{equation}
By the law of large numbers,  $\P(\hat{s}_1 = \hat{s}_2) \rightarrow 1$ as $M_1 \rightarrow \infty$. Hence $D\left(P(\hat{s}_1,\hat{s}_2) || P(\hat{s}_1)P(\hat{s}_2)\right) \rightarrow H(S)$ and $D\left(P(\hat{s}_1) || Q(\hat{s}_1) \right) \to D\left(P(s) || Q(s) \right)$ as $M_1 \rightarrow \infty$. Therefore, we can find a lower bound on the exponent of the probability of error by solving the following optimization problem:
\begin{equation}
E_2= \min_{P\in \PP} \sum_{s\in\{A,C,G,T\}} P(s) \log \left( \frac{P(s)}{Q(s)^2} \right).
\end{equation} 
One can show that the preceding optimization problem is minimized with $P(s) = \frac{Q(s)}{\sum_{s} Q(s)^2}$. Therefore, 
\begin{equation}
E_2= \Hr.
\end{equation}
Since $\L > 2/\Hr$, $\P(\EE_2(M_1)) \rightarrow 0$ as $M_1,G \rightarrow \infty$.

\paragraph{ $\P(\EE_3(M-\sqrt{M})) \rightarrow 0$}
Fix a particular alignment , say $\bU$, with $M-\sqrt{M}$ reads from same location.  Let  us call the DNA subsequence at that location the source sequence. A simple large deviations argument says that the probability that the ML estimate is not the same as the corresponding base of the source sequence is  bounded by $2^{-M\gamma}$ for some $\gamma > 0$. Hence, the probability that the averaged sequence from $\bU$ is at distance greater than $K\tau_G$  from the source sequence is bounded by
$$  2^{-KD(\tau_G || 2^{-M\gamma})}.$$
For large $M$, this is approximately $2^{-KM\gamma \tau_G}$. By the union bound,
\begin{equation}
\label{eq:e3}
Pr(\EE_3(M-\sqrt{M}))) < A 2^{-KM\gamma \tau_G},
\end{equation}
where $A$ is the number of such alignments.  Let us bound $A$. First, the $\sqrt{M}$ K-mers from other locations  can at most come from $G^{\sqrt{M}}$ different locations.  Let us now look at the number of possible choices of the $M -\sqrt{M}$ $K$-mers coming from the same location. There are $G$ possible such locations. It is easy to show that for some constant $\eta > 0$, with high probability at each location there are at most $\eta \log G$ K-mers.  Hence, the number of choices of the $M-\sqrt{M}$ $K$-mers coming from the same location is bounded by $G(\eta \log G)^{M-\sqrt{M}}$. Hence,
$$ A < G^{\sqrt{M}} \times G(\eta \log G)^{M-\sqrt{M}}, $$
and from (\ref{eq:e3}), we get:
$$Pr(\EE_3(M-\sqrt{M}))) <  G^{\sqrt{M}+1}(\eta \log G)^{M-\sqrt{M}} 2^{-KM\gamma \tau_G}.$$
Recall that $M = \beta \log G$, $ K = \L \log G - (\L \log G)^\alpha$.  A direct calculation shows that $\P(\EE_3(M - \sqrt{M}))$ can be driven to zero if we choose $\tau_G = (\log G)^{-\frac{1}{4}} \rightarrow 0$, for example.

%
%
%
%
%
%
%
%
%
%
%
%
%
%
%
%
%
%
%
%
%
%

\subsubsection{Coverage Condition}
To prove the coverage condition of Theorem \ref{thm:main}, we need to show that $\tilde{\RR}$ contains enough cleaned-up $K$-mers covering the DNA sequence. We say a substring of length $K$ is covered if there exists a cleaned up read $\bar \br$ which is within  Hamming distance $K\tau$ of the substring. Similarly, we say a base is covered if one of the substrings containing the bases is covered by a member of $\tilde{\RR}$.  

Let $\CC$ be the event that there exists a base not covered by members of $\tilde{\RR}$. Let $\CC_i$ be the event that the $i$th base is not covered. Clearly, $\CC= \cup_{i=1}^{G} \CC_i$. Using the union bound and considering the fact that coverage condition is symmetrical for all bases, we obtain
\begin{equation}
\P(\CC) \leq G \P(\CC_K).
\end{equation}
There are $K$ substrings of length $K$ containing the $K$th base of the DNA sequence. If none of the substrings is covered, then the event $\CC_K$ happens. 

Let $f$ denote the probability that a given substring of length $K$ in the DNA sequence is not covered due to the reads containing it. We claim that
\begin{equation}
\P(\CC_K) \leq f^{\frac{K}{L-K}}.
\end{equation}
To see this, instead of considering all the substrings covering the $K$th base, we consider only a subset of them consisting of $\frac{K}{L-K}$ substrings with starting positions at $(L-K)j$ for $j\in \{1,\dots, \frac{K}{L-K} \}$. For each substring, the probability of not being covered is at most $f$ and there exists no read that can contain two of the substrings. Therefore, the probability of missing all the substrings becomes $f^{\frac{K}{L-K}}$ due to independence of probabilities. The inequality comes from the fact that there are cases where the $K$th base is covered but not from the chosen subset of substrings.

We need to obtain an upper bound on $f$. We look at the interval of length $L-K$ before the given substring. The number of reads $k$ with starting location in the interval has Poisson distribution with parameter $\lambda (L-K)$. We assume that no other read sampled from outside of the interval can assist us in cleaning-up the substring. Clearly, if $0\leq k<M$ then there is not enough reads and hence the substring is not in $\tilde{\RR}$. For $iM \leq k < (i+1)M$, we partition the reads into $i$ disjoint sets each of which having $M$ members. For this case, the substring is not covered if  none of the sets is a good alignment. For given $i$, let $\DD_j$ for $j\in \{1,\dots,i\}$ be the event that the $j$th subset is not a good alignment. The probability that the substring is not covered is $\P(\cap_{j=1}^i \DD_j)$. We claim that the events $\DD_j$ are independent and therefore,
$\P(\cap_{j=1}^i \DD_j)  = \prod_{j=1}^i \P(\DD_j)= \left(\P(\DD_1)\right)^i$. This is due to the fact that the type of the substring irrespective of being close to $Q$ or not is included in the definition of ``good alignment''. In fact, the definition of good alignment is universal and includes all mother substrings present in the DNA sequence. 

Using large deviation arguments, the probability  that one of the sets fails to pass the good alignment test is bounded by $2^{- K\zeta(\epsilon)}$, for some $\zeta(\epsilon)> 0$. Therefore,

\begin{align*}
f \leq \sum_{k=0}^{\infty} \frac{ (\lambda (L-K))^k e^{-\lambda (L-K)}}{k!} 2^ {\lfloor k/M \rfloor K \zeta(\epsilon)}.
\end{align*}
We can upper bound it further by
$f  \leq e^{-\lambda (L-K) (1-2^{- \frac{K\zeta(\epsilon)}{M}})}.$
Using the upper bound  on $f$, we obtain
\begin{equation}
\P(\CC) \leq G e^{-\lambda K \left(1-2^{- \frac{K\zeta(\epsilon)}{M}}\right)}.
\end{equation}
One can show that if $N> \frac{G}{L}\ln(G)=N_{\text{cov}}$ then $\P(\CC) \to 0$.


\section*{Acknowledgements}
This work is supported by the Natural Sciences and Engineering
Research Council (NSERC) of Canada  and by the Center for Science of Information (CSoI), an NSF Science and Technology Center, under grant agreement CCF-0939370.

\bibliographystyle{amsplain}
\bibliography{references}

\end{document}